\documentclass[prl,preprintnumbers,twocolumn,nofootinbib]{revtex4-1}
\usepackage{amssymb,amsmath,amsthm}
\usepackage[dvips]{graphicx}
\usepackage{verbatim}
\usepackage{epsfig}
\usepackage{color}
\usepackage{comment}
\usepackage{subfigure}

\usepackage{graphicx}
\usepackage{bm}  
\usepackage{epsfig}
\usepackage{mathtools}
\usepackage{verbatim} 

\begin{document}

\title{Splay--density coupling in semiflexible main-chain nematic polymers with hairpins}

\date{\today}

\author{ Aleksandar Popadi\' c$^{1}$, Daniel Sven\v sek$^{2,*}$, Rudolf Podgornik$^{2,3}$, Kostas Ch. Daoulas$^4$, and Matej Praprotnik$^{1,2}$}

\affiliation{
\mbox{$^1$Laboratory for Molecular Modeling, National Institute of Chemistry, SI-1001 Ljubljana, Slovenia}\\
\mbox{$^2$Department of Physics, Faculty of Mathematics and Physics, University of Ljubljana, SI-1000 Ljubljana, Slovenia}\\
\mbox{$^3$Department of Theoretical Physics, J. Stefan Institute, SI-1000 Ljubljana, Slovenia}\\
\mbox{$^4$Max Planck Institute for Polymer Research, 55021 Mainz, Germany}\\
$^{*}$email: daniel.svensek@fmf.uni-lj.si}

\begin{abstract}
\noindent 
We establish a macroscopic description of the splay--density coupling in semiflexible main-chain nematic polymers with hairpins, using a vectorial continuity constraint for the ``recovered'' polar order of the chain tangents and introducing chain backfolds (hairpins) as its new type of sources besides chain ends. We treat both types of sources on a unified basis as a mixture of two ideal gases with fixed composition. Performing detailed Monte Carlo simulations of nematic monodomain melts of ``soft'' worm-like chains with variable length and flexibility, we show via their structure factors that the chain backfolding weakens the splay--density coupling, and demonstrate how this weakening can be consistently quantified on the macroscopic level. We also probe and discuss the deviations from the noninteracting gas idealization of the chain ends and backfolds.

\end{abstract}
\maketitle


\noindent
Formal description of {\it line liquids} \cite{KamienNelson,nelson,Nelson} differs fundamentally from hydrodynamic description of isotropic and ordinary nematic liquids since the connectivity of the 
oriented lines stipulates an additional explicit macroscopic constraint \cite{degennes,meyer_chapter}. This is true for equilibrium or living main-chain polymers, self-assembled molecular chains as well as worm-like micelles, whose consistent description implies a {\sl conservation law} stemming directly from their unbroken connectivity. The exact nature and form of this conservation law proposed independently by de Gennes and Meyer, received recently a renewed scrutiny \cite{svensek_tensorial} that uncovered its deeper structure and important consequences missed in the previous analysis. In fact, 
its consequences
trickle all the way down
to fundamental macroscopic, observable properties such as structure factors and coarse-grained order parameters \cite{svensek_jcp} as in, e.g., the ordered and/or confined phases of DNA \cite{Shin,svensek-podgornik,julija}. The fundamental issue that we address in this contribution is the way this conservation law enters the coarse-grained Ornstein-Zernicke free energy description of a nematic polymer with arbitrary chain backfolding and, specifically, the magnitude of the corresponding phenomenological coupling strength \cite{kamien}. By comparing detailed simulations based on a recently developed mesoscopic model~\cite{kostas_soft} with the predictions of the Ornstein-Zernicke description augmented by the conservation law, we derive an explicit form of the coupling strength that takes into account the nematic order as well as the hairpin folds.

It has been recognized a while ago \cite{degennes,meyer_chapter,odijk88,KamienNelson,meyer,selinger,nelson,nelson0,kamien,selingerJP} that the connectivity of the polymer chain manifests itself on the macroscopic level as a constraint on the continuum fields (i.e., order parameter and density/concentration) describing the coarse-grained version of the polymer configuration.  For the nematic director field ${\bf n}({\bf r})$, such constraint was written in form of a conservation law \cite{degennes}
\begin{equation}
	\nabla\cdot (\rho_s {\bf n}) = \rho^+ - \rho^-,
	\label{vectorial_s}
\end{equation}
where $\rho_s({\bf r})$ is the surface density or concentration of polymer chains perforating the plane perpendicular to ${\bf n}({\bf r})$ and $\rho^+({\bf r})$ and $\rho^-({\bf r})$ are volume densities of the beginnings and endings of chains acting as sources in this continuity equation for the ``polymer current'' $\rho_s {\bf n}$.

As shown recently within a more formal framework \cite{svensek_tensorial}, Eq.~(\ref{vectorial_s}) generalizes to a continuity equation for the full order vector ${\bf a}({\bf r})$,
%
%
\begin{equation}
	\nabla\cdot(\rho\ell_0{\bf a}) = \rho^+ - \rho^-,
	\label{vectorial}
\end{equation}
where $\rho({\bf r})$ is now the volume number density of arbitrary segments (e.g. monomers) of length $\ell_0$; with that, $\rho_s = \rho\ell_0 |{\bf a}|$. 
Eq.~(\ref{vectorial}) represents a conservation law for the polymer current ${\bf j} = \rho\ell_0{\bf a}$, where it is clear by construction \cite{svensek_tensorial} that ${\bf a}({\bf r}) = \langle {\bf t}\rangle$ is exactly the {\it polar} order of polymer chain tangents ${\bf t}$. 
Eq.~(\ref{vectorial_s}) is a special case of Eq.~(\ref{vectorial}) for $|{\bf a}|={\rm const.}$ and is therefore of the same, polar type, where $\bf n$ cannot be anything but a polar(!) preferred direction.

Notwithstanding the inconvenient fact that nematic ordering is apolar and does not possess a polar quantity like $\bf a$, the constraint Eq.~(\ref{vectorial_s}) has been readily applied to main-chain nematic polymers. Moreover, the behavior of macroscopic observables in recent simulations of polymer nematics \cite{kostas_soft} was consistent with Eq.~(\ref{vectorial_s}).
Usually, the argument that hairpins (sharp, ideally point-like 180$^\circ$ turns of the chain) be absent has been invoked to circumvent the problem of the vanishing order vector and validate the use of the director $\bf n$, while it has been at the same time recognized theoretically \cite{odijk88,kamien,terentjev} that hairpins act as chain ends and their density defines an effective length of the chains.

Real-space symmetric tensor (quadrupolar) fields, like the undirected axis described by $\bf n$, are less frequent in nature than vector or scalar fields. They appear mainly in the context of orientational ordering (nematic liquid crystals, natural patterns of various kinds), while continuity equations for quadrupolar fields are rarely encountered.
Here, we present a well-grounded example of how one can substitute the conservation law for a quadrupolar nematic variable with a more usual conservation of a vector current, by restricting the manifold of the latter to the projective plane.

\begin{figure}[h]
\begin{center}	
	\includegraphics[height=36mm]{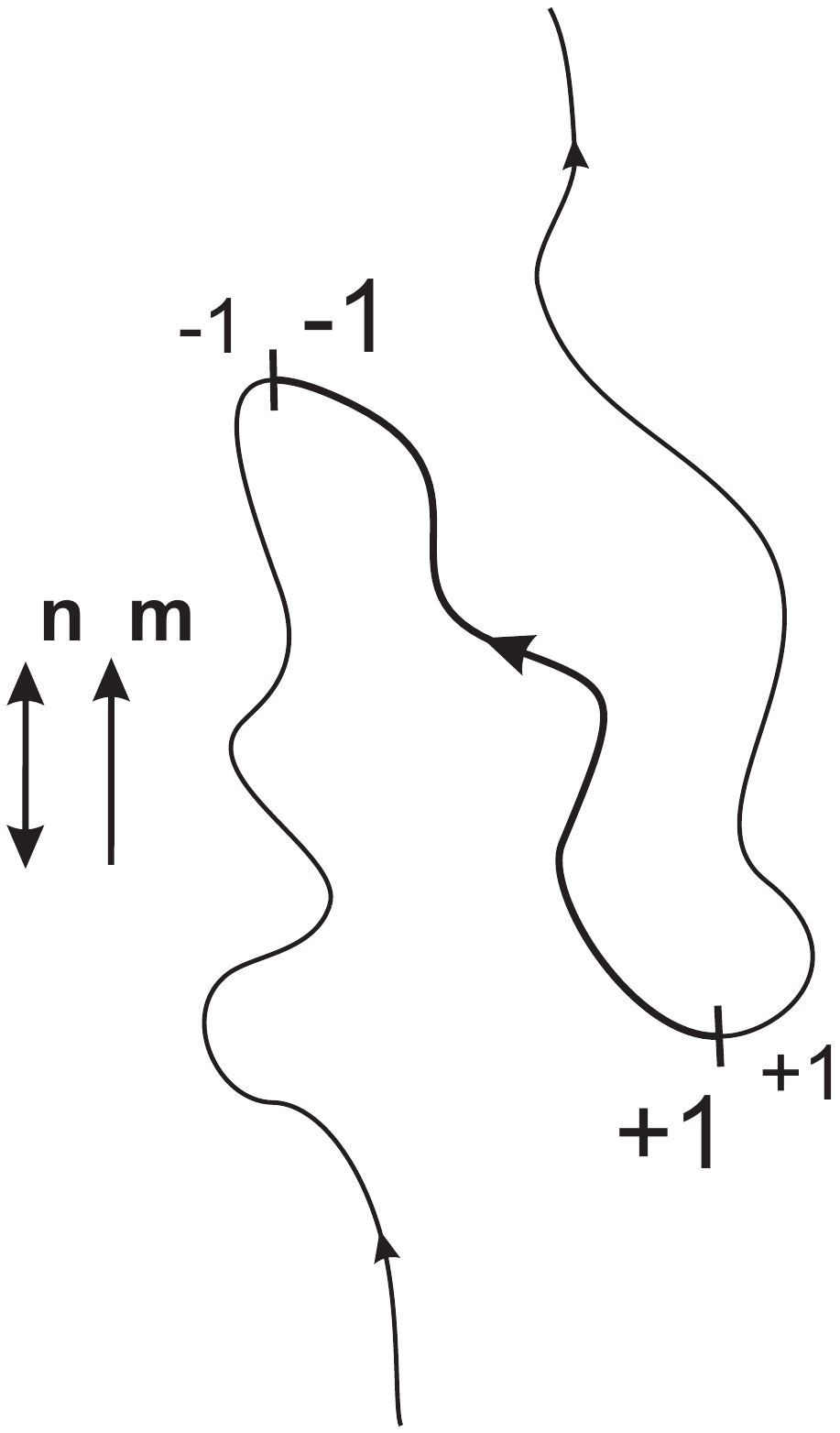}
	\includegraphics[height=35.7mm]{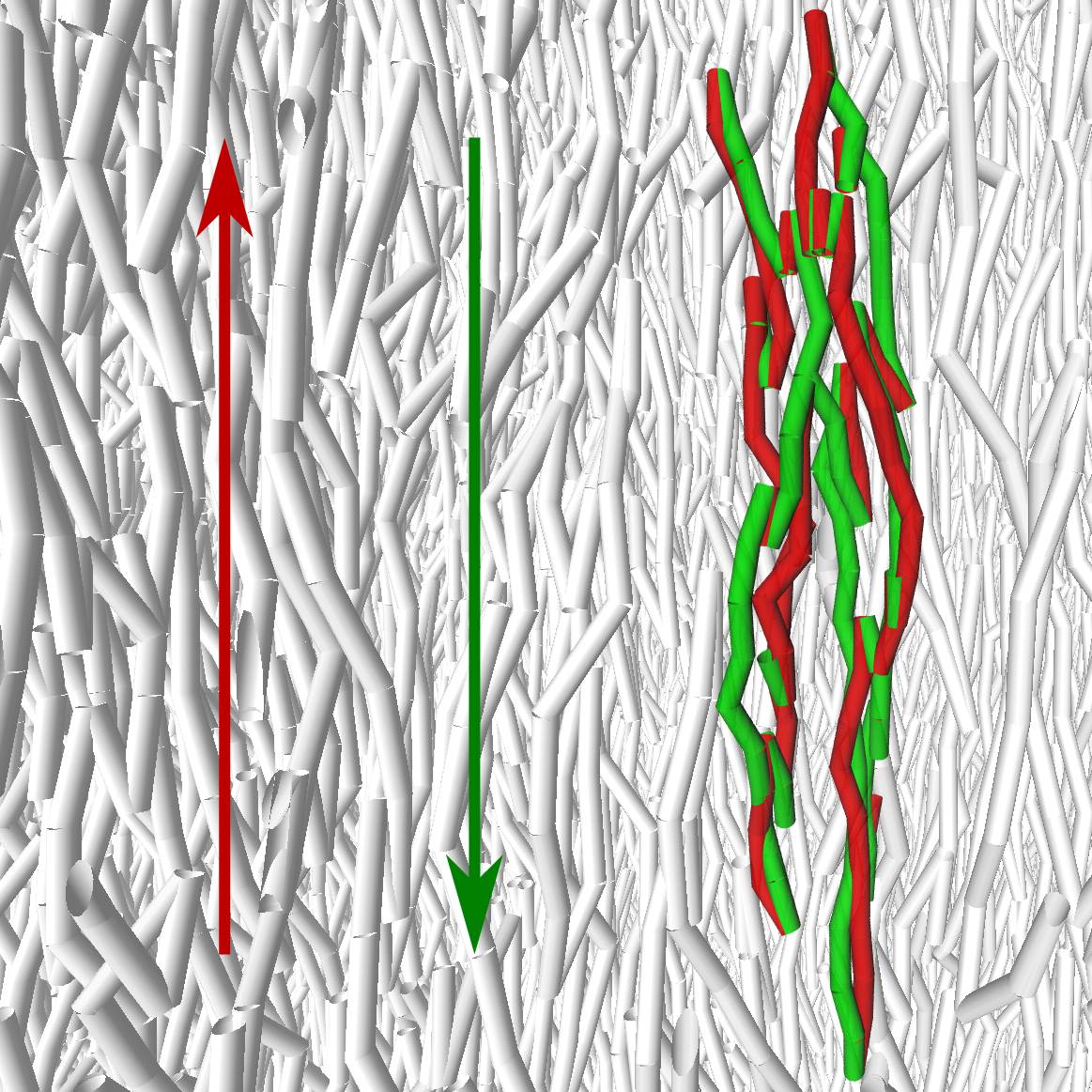}
	\includegraphics[height=37mm]{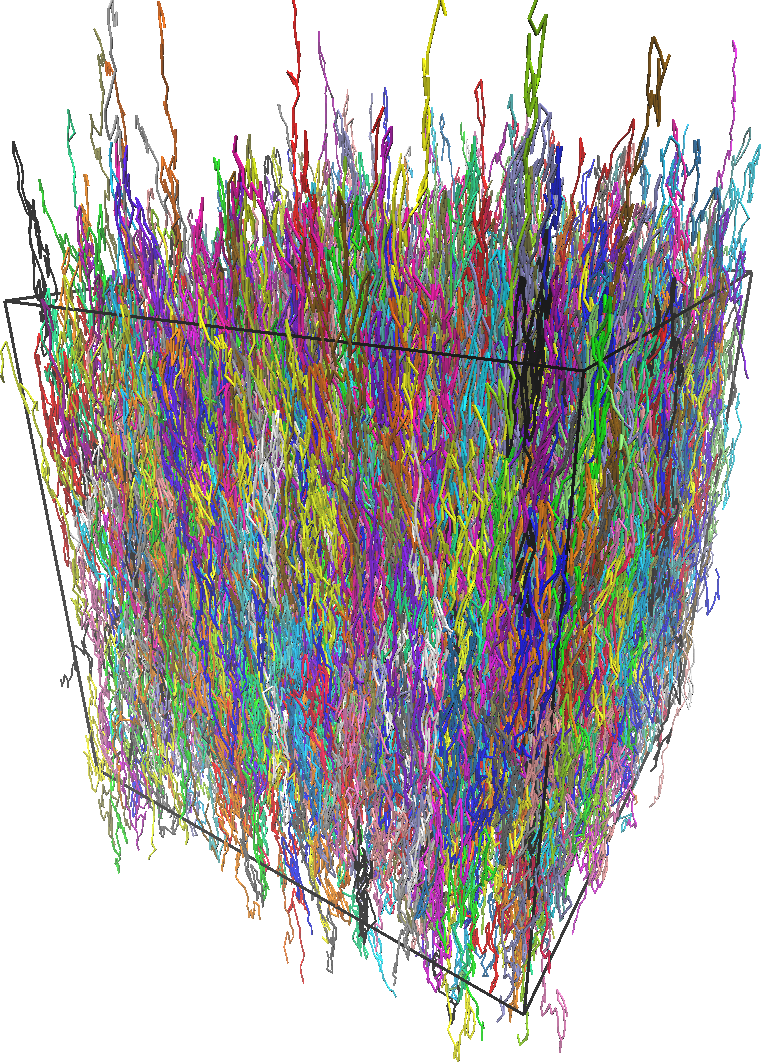} 
	\caption{Left: pair of virtual cuts at points of folding (${\bf t}\cdot{\bf m}=0$) with respect to a chosen polar direction ${\bf m}\parallel{\bf n}$ and inversion of the backfolded (${\bf t}\cdot{\bf m}<0$) segments introduce a pair of separated $+2$ source and $-2$ sink, while thus-emerging polar order is independent of the folding. Middle: example of a single chain with folds (hairpins), belonging to the simulated melt (right) with $2^{18}$ monomers.}
\label{fig:backfoldings}
\end{center}
\end{figure}

Recently, it has been indicated \cite{svensek_generalized} that the vectorial conservation Eq.~(\ref{vectorial}) could be consistently applied to a nematic polymer with arbitrary number of hairpins or finite-size backfolds by introducing a so-called ``recovered polar order'' ${\bf a}^r({\bf r})\parallel {\bf n}({\bf r})$ of chain tangents and accompanying additional chain beginnings and endings of strength $\pm 2$ corresponding to virtual backfolding cuts, Fig.~\ref{fig:backfoldings} (left). A rigorous conservation law
for the recovered polar order is then
\begin{equation}
	\nabla\cdot(\rho\ell_0 {\bf a}^r) = \Delta\rho^{s\pm},
    \label{recovered}
\end{equation}
where the source $\Delta\rho^{s\pm} = \rho^+ - \rho^- + 2\rho^{2+} - 2\rho^{2-}$, besides mismatching physical chain ends, now contains also a contribution from mismatching densities $\rho^{2+}$ and $\rho^{2-}$ of up and down chain backfolding virtual cuts.
Here we present the first direct evidence for the relevance of this suggestion, employing extensive Monte Carlo (MC) simulations of a ``soft'' model of worm-like chains (WLCs) \cite{kostas_soft} and tracing the signal of the constraint Eq.~(\ref{recovered}) expressed in terms of the recovered polar order ${\bf a}^r$.
With that, we show that the semiflexibility of the polymer chain can be consistently taken into account on the macroscopic level and  hairpins can be rigorously incorporated as sources in the continuity constraint on the macroscopic fields.



In the following two paragraphs, we prepare the prerequisites needed for the leanest possible description of the sources, which brings about a minimum number of additional parameters and does not introduce any additional variables. Such first-step minimalism is an intentional convenience and does not mean that subsequent extensions and refinements are ruled out.

Since in an apolar system there is no distinction between chain beginnings and endings ($\equiv$ chain ends), we can 
without loss of generality 
consider the deviations of the densities of both of these chain end types from the equilibrium value $\rho_0^+ = \rho_0^- \equiv {\textstyle{1\over 2}}\rho_0^\pm$ to be symmetric\footnote{This is a trivial statement. In a nematic, there can be no physical distinction between beginnings and endings. Moreover, the arbitrary choice of the direction of chain parametrization cannot influence any physical configuration whatsoever: selecting at random a chain end anywhere in the system, under {\it any condition}, there is no preference towards the beginning or ending. In a nematic, $\rho^+$ and $\rho^-$ are not separate variables. There is only one variable, $\Delta\rho^{\pm}$.}, 
\begin{eqnarray}
	\rho^+ - \rho_0^+ = -(\rho^- - \rho_0^-) &\equiv & {\textstyle{1\over 2}} \Delta\rho^\pm,\label{Deltarho}\\
	\rho^{2+} - \rho_0^{2+} = -(\rho^{2-} - \rho_0^{2-}) &\equiv & {\textstyle{1\over 2}} \Delta\rho^{2\pm},\label{Deltarho2}
\end{eqnarray}
where an analogous statement, Eq.~(\ref{Deltarho2}), holds for up and down backfolds with the equilibrium density $\rho_0^{2+} = \rho_0^{2-} \equiv {\textstyle{1\over 2}}\rho_0^{2\pm}$.
This does not imply in any way that the number of backfolds per chain must be even. It just reflects the symmetry-based facts that i) in a homogeneous equilibrium system the number of up and down backfolds is equal on average, that ii) the deviations $\rho^{2+} - \rho_0^{2+}$ and $\rho^{2-} - \rho_0^{2-}$ from the homogeneous distributions are equally costly, iii) that in the source of Eq.~(\ref{recovered}) $\rho^{2+} - \rho_0^{2+}$ is equivalent to $-(\rho^{2-} - \rho_0^{2-})$ (excess of up backfoldings has the same effect as shortfall of down backfoldings), and iv) that we will not distinguish between these two types of sources. As long as $\Delta\rho^{2\pm}$ is much smaller than $\rho_0^{2\pm}$, this distinction plays no role.

Considering
the chain ends and the backfolding cuts as two types of free noninteracting particles (two ideal gases), the free-energy cost of their nonequilibrium distribution is entropic \cite{nelson},
\begin{equation}
	f(\Delta\rho^\pm, \Delta\rho^{2\pm}) = {k_{\rm B} T\over 2}\left[
    	{(\Delta\rho^\pm)^2\over\rho_0^\pm} + {(\Delta\rho^{2\pm})^2\over\rho_0^{2\pm}}
        \right].
    \label{fboth}
\end{equation}
Moreover, we want to treat both types of particles on an equal basis and describe the source in Eq.~(\ref{recovered}) by the single variable 
\begin{equation}
	\Delta\rho^{s\pm} = \Delta\rho^\pm + 2\Delta\rho^{\pm 2},
    \label{total}
\end{equation} 
without considering its breakdown into the two individual contributions. The free-energy density of the total source $\Delta\rho^{s\pm}$ is then obtained by averaging Eq.~(\ref{fboth})
over all possible realizations Eq.~(\ref{total}) of $\Delta\rho^{s\pm}$:
\begin{equation}
	\bar{f}(\Delta\rho^{s\pm}) = \int_{-\infty}^{\infty}\!\!\!\!\!\!\!\!{\rm d}\Delta\rho^{\pm}
	f(\Delta\rho^\pm, \Delta\rho^{2\pm})
	{\cal P}(\Delta\rho^{\pm}){\cal P}(\Delta\rho^{2\pm}),
    \label{fbar}
\end{equation}
with $\Delta\rho^{2\pm} = (\Delta\rho^{s\pm}-\Delta\rho^\pm)/p \equiv \Delta\rho^{p\pm}$ and $p$(=2) introduced for trackability, where ${\cal P}(\Delta\rho^{\pm})$  and ${\cal P}(\Delta\rho^{p\pm})$ are Gaussians with variance $\rho_0^{\pm}/V_1$ and $\rho_0^{p\pm}/V_1$, respectively,
$V_1$ is an arbitrary volume (e.g.~the coarse-graining volume) not appearing in the final result and the normalization is 
$\int_{-\infty}^{\infty}\!\!\!{\rm d}\Delta\rho^{\pm}\,\,{\cal P}(\Delta\rho^{\pm})\,{\cal P}(\Delta\rho^{p\pm}) = 1$.
Omitting a constant term $k_{\rm B} T/(2 V_1)$ (arising due to the fact that the state $\Delta\rho^{s\pm}=0$ can be realized by $\Delta\rho^{\pm}=-p \Delta\rho^{p\pm}\ne 0$, which costs energy), 
the result of Eq.~(\ref{fbar}) is the average free-energy density of the total source $\Delta\rho^{s\pm}$,
\begin{equation}
	\bar{f}(\Delta\rho^{s\pm}) = {k_{\rm B} T\over 2}\, {(\Delta\rho^{s\pm})^2\over \rho_0^{\pm}+4 \rho_0^{2\pm}}
    \equiv	{1\over 2} G (\Delta\rho^{s\pm})^2,
    \label{f_avg}
\end{equation}
which presents a penalty potential with strength $G$ of the 
constraint Eq.~(\ref{recovered}) for the recovered polar order.

Neglecting variations of nematic order modulus and expanding $\rho = \rho_0+\delta\rho({\bf q})$, ${\bf n} = {\bf n}_0+\delta{\bf n}({\bf q})$ around equilibrium values $\rho_0$, ${\bf n}_0$, the free-energy \cite{nelson,svensek_jcp} contribution of a Fourier mode ${\bf q}=({\bf q}_\perp, q_\parallel)$ in the volume $V$ is $F_{\bf q} = f({\bf q})/V$ and 
\begin{eqnarray}
	f({\bf q}) &=& {1\over 2}\tilde G\left|q_\parallel{\delta\rho\over\rho_0}+q_\perp \delta n_L\right|^2
    				+{1\over 2}B\left|\delta\rho\over\rho_0\right|^2  \label{f_q} \\ 
                        &+& 
							{1\over 2}(K_1 q_\perp^2 + K_3 q_\parallel^2) |\delta n_L|^2 + 
							{1\over 2}(K_2 q_\perp^2 + K_3 q_\parallel^2) |\delta n_T|^2,\nonumber
\end{eqnarray}
where ${\bf n}_0\cdot\delta{\bf n}=0$, $q_\parallel$ is the component along ${\bf n}_0$
and $\delta n_L$, $\delta n_T$ are the longitudinal and transverse components with respect to ${\bf q}_\perp$, $B$ is the compressibility modulus and $K_{\{1,2,3\}}$ are the \{splay,twist,bend\} elastic constants.
The first term of Eq.~(\ref{f_q}) is exactly the free-energy cost Eq.~(\ref{f_avg}) of the total source $\Delta\rho^{s\pm}$, expressed by the left-hand-side of Eq.~(\ref{recovered}); here
$\tilde G = G (\rho_0 \ell_0 a_0^r)^2$ and $a_0^r$ is the magnitude of the recovered polar order.

\begin{figure}[h]
\begin{center}	
	\mbox{
    \includegraphics[height=36mm]{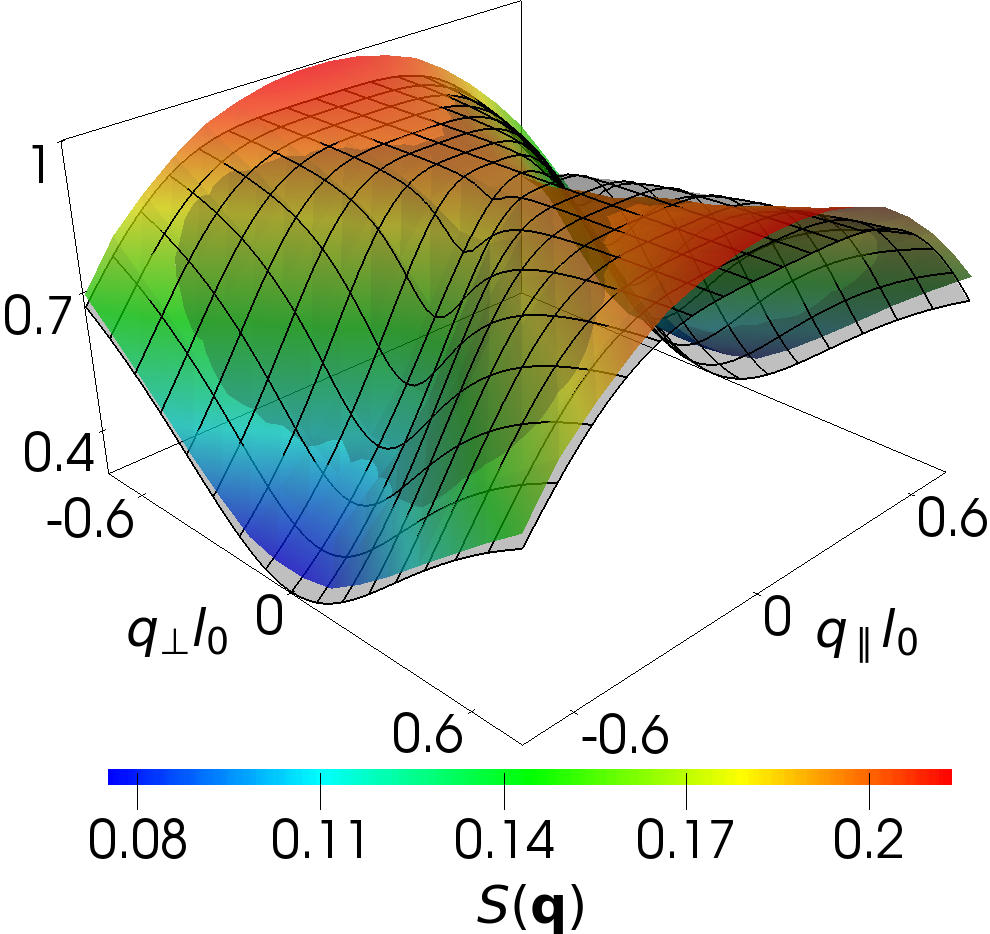}
    \includegraphics[height=36mm]{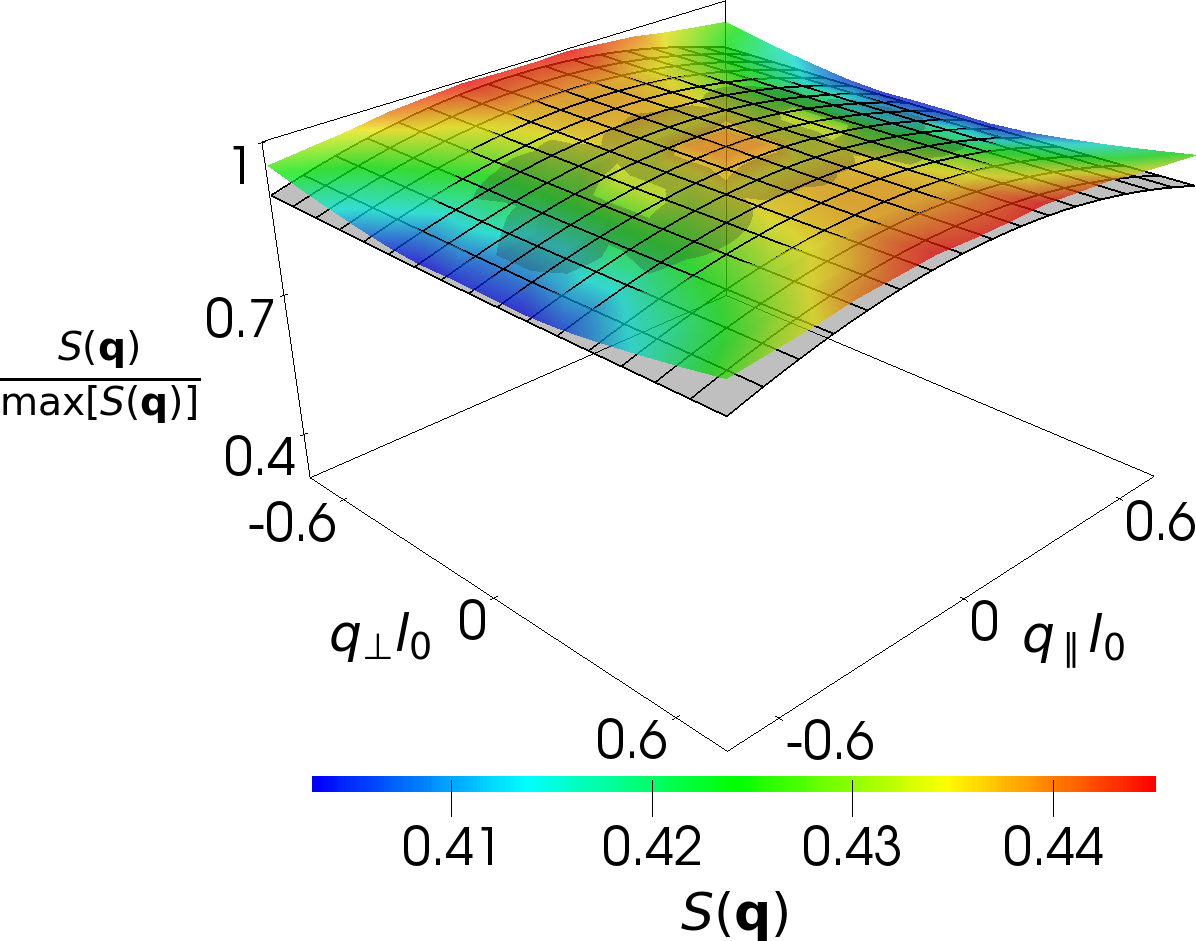}
    }
	\mbox{
    \includegraphics[height=34mm]{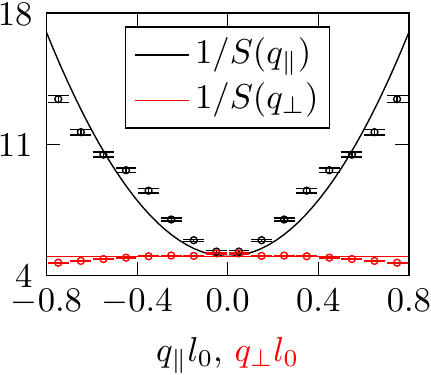}
    \includegraphics[height=34mm]{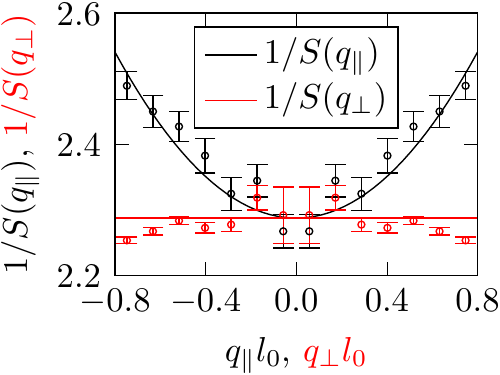}
    }
	\caption{Top: structure factors $S({q}_\perp l_0,q_\parallel l_0)$ calculated in simulations (solid), scaled to their maximum values and fitted (wireframe) by Eq.~(\ref{Svect}), for stiff (left column) and flexible (right column) chains of length $128l_0$; $l_0$ is the length of the WLC segment. Bottom: cross sections of $S^{-1}$ for $q_\perp\!=\!0$ (black) and $q_\parallel\!=\!0$ (red).}
\label{fig:S(q)}
\end{center}
\end{figure}

The structure factor $S({\bf q}) = \langle \delta\rho({\bf q})\delta\rho(-{\bf q})\rangle/N$ (where $N$ is the total number of $\ell_0$ segments) corresponding to 
Eq.~(\ref{f_q}) is then found to be \cite{nelson,svensek_jcp}
\begin{equation}
	S({\bf q}) = 
	k_{\rm B} T \rho_0{q_\perp^2 + {\left(K_1 q_\perp^2+K_3 q_\parallel^2\right)/\tilde{G}}\over
					 {B} q_\perp^2 + \left({{B}/\tilde{G}}+q_\parallel^2\right)\left(K_1 q_\perp^2+K_3 q_\parallel^2\right)}.
	\label{Svect}
\end{equation}
Its dependence on $\tilde G$ makes it a suitable signal for detecting the constraint Eq.~(\ref{recovered}) and determining its strength from molecular-simulation data, Fig.~\ref{fig:S(q)}.

Verifying the predictions of the macroscopic theory with molecular-level computer simulations of polymer
nematics is challenging, since such simulations must i) address the long-wavelength limit and ii) realize different regimes of chain backfolding (hairpin formation). Thus, it is essential to consider large systems containing long polymer chains \cite{kremer-grest}. We fulfill these requirements benefiting from a recently developed mesoscopic model \cite{kostas_soft} which describes the polymers as discrete WLCs, Fig.~\ref{fig:backfoldings} (middle, right). 
The modeled system contains $N_{\rm c}$ WLCs comprised of $N_{\rm s}$ linearly connected segments of fixed length $l_0$. Consecutive
segments are subjected to a standard angular potential $U_{\rm b} = -\epsilon {\bf u}^{\rm i}(s)\cdot {\bf u}^{\rm i}(s+1)$,
where ${\bf u}^{\rm i}(s)$ is the unit vector along the $s$-th segment of the $\rm i$-th chain and
$\epsilon$ controls the WLC stiffness. Non-bonded interactions between segments are introduced via
the potential $U_{\rm nb} = U(r^{\rm ij}(s,t))\left[\kappa - (2\upsilon/3){\bf q}^{\rm i}(s)\;{\bf :}\;{\bf q}^{\rm j}(t)\right]$, where
$U(r^{\rm ij}(s,t)) = C_{0}\Theta \left(2\sigma - r^{\rm ij}(s,t)\right)\left[4\sigma+r^{\rm ij}(s,t)\right]\left[2\sigma-r^{\rm ij}(s,t)\right]^2$ 
and $r^{\rm ij}(s,t)$ is the distance between the centers of the $s$-th and $t$-th segments of the $\rm i$-th and $\rm j$-th chain, respectively.
The interaction range is controlled by $\sigma$ as indicated by the Heaviside function $\Theta$. To verify the predictions of the macroscopic theory
it is sufficient to employ a generic model with a single ``microscopic'' length scale. Hence, we set $\sigma = l_0$, although other choices are
possible~\cite{daoulas,greco} when modeling actual materials. The integrated strength of $U(r^{\rm ij}(s,t))$ is normalized to $l_0^3$, choosing
$C_{0} = 3l_0^3/(64\pi \sigma^6)$. The strength of the isotropic repulsion between the segments is controlled by the parameter $\kappa$. Nematic alignment
is promoted by the anisotropic part of $U_{\rm nb}$, which depends on the 
inner product of tensors ${\bf q}^{\rm i}(s) = \left[3{\bf u}^{\rm i}(s) \otimes {\bf u}^{\rm i}(s) - {\bf I}\right]/2$
quantifying the segmental orientation in the laboratory frame. The strength of these Maier-Saupe-like interactions is controlled by $\upsilon$.

Two molecular flexibilities $\epsilon = 0$ and $\epsilon=3.284k_{\rm B} T$ are addressed, corresponding to flexible and stiff chains, respectively. In both cases, we consider WLCs with $N_{\rm s} = \{32,64,128\}$ segments. We empirically set $\kappa = 7.58k_{\rm B} T$, while $\upsilon = 3.33k_{\rm B} T$ and $6.66k_{\rm B} T$ for the stiff and flexible chains, respectively. For this $\kappa$, the repulsive interactions are strong enough to furnish a stable polymer liquid (positive compressibility~\cite{daoulas}) but remain
sufficiently ``soft'' for efficient simulations. Our choices of $\upsilon$ lead to stable nematic order in 
all considered cases.

We study large nematic monodomains containing 
$N = N_{\rm c}N_{\rm s} = 2^{18}$ segments, Fig.~\ref{fig:backfoldings} (right). 
They
are equilibrated through MC starting from configurations where all chains are stretched and aligned along the $z$-axis of the laboratory frame, having their
centers-of-mass randomly distributed. The MC algorithm utilizes standard~\cite{Frenkel,Tuckerman} slithering-snake moves, as well as volume fluctuation moves at
pressure $Pl_0^3/(k_{B}T) = 2.87$ resulting in system's volume fluctuations of $\sim 1\%$. Working in the isothermal-isobaric ensemble is computationally more expensive, 
however preferred here to exclude isotropic/nematic coexistence in the entire range of considered parameters. The efficient soft model enables us to accumulate large sequences of nematic melt monodomain configurations, which allow for direct verification of the macroscopic theory via the structure factor 
Eq.~(\ref{Svect}) as follows.

The global nematic direction and the ensemble volume are free to fluctuate. Therefore we compute the structure factor $S(q_x, q_y, q_z)$ of each configuration in the laboratory frame and assign it to a bin representing $S({q}_\bot, q_\parallel)$, where $q_\parallel$ and ${q}_\perp=|{\bf q}_\perp|$ are the components parallel and orthogonal to the current nematic director determined as the principal eigenvector of $(1/N)\sum_{\mathrm{i}, s} \mathbf{q}^\mathrm{i}(s)$. With that, $S({q}_\bot, q_\parallel)$ is computed in the director-based ${\rm 123}$ frame \cite{Allen1996,Masters2008,kostas_soft}.
The principal eigenvector is also used to determine, for each configuration, the modulus of the recovered polar order $a_0^r$ appearing in the definition of $\tilde G$, which is then averaged over all recorded configurations. The same is done for the density of segments $\rho_0=N/\langle V\rangle$ (putting $\ell_0=l_0$) in Eq.~(\ref{Svect}), as well as the densities of chain ends $\rho_0^\pm$ and backfolds $\rho_0^{2\pm}$ in Eq.~(\ref{f_avg}). 
In all cases, block-averaging with block size $\tau$ is employed, where $\tau$ is the number of MC steps needed to decorrelate the end-to-end vector of the WLC. Computationally most severe are stiff chains with $N_{\rm s}=128$ segments, where $\tau$ was as high as 130\,000 and a MC sequence of $48\tau$ was reached. In other cases the runs in terms of $\tau$ were longer.

\begin{figure}[h]
\begin{center}	
    \includegraphics[height=42.5mm]{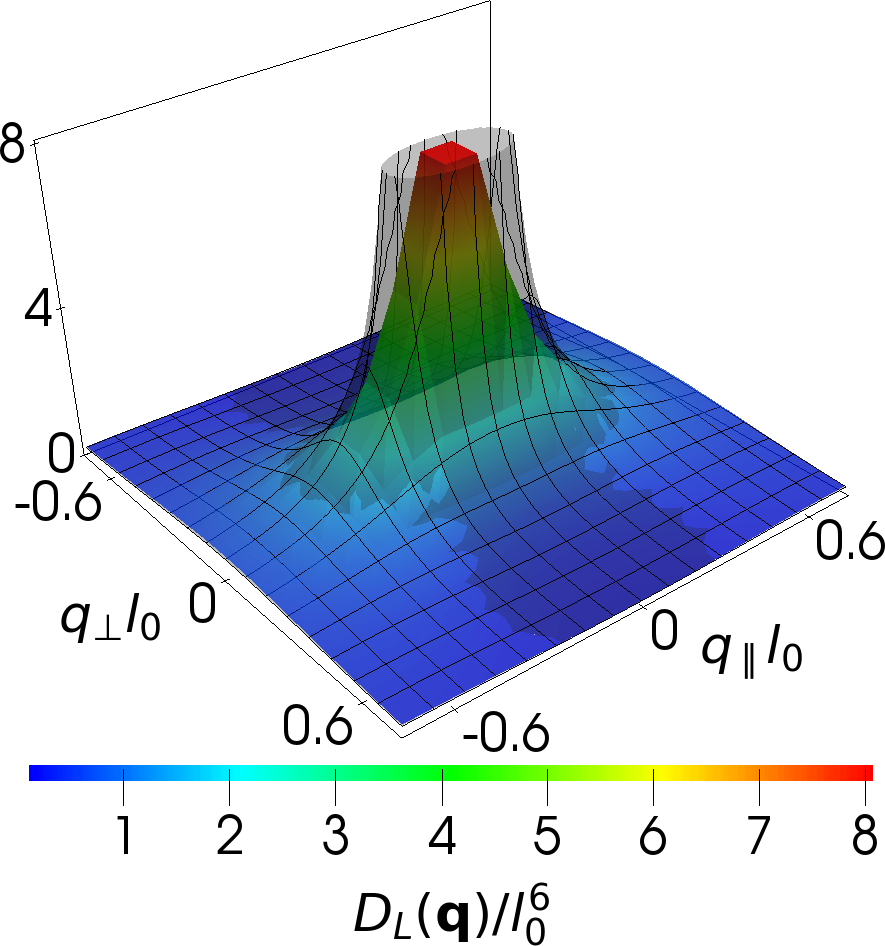}
    \includegraphics[height=42.5mm]{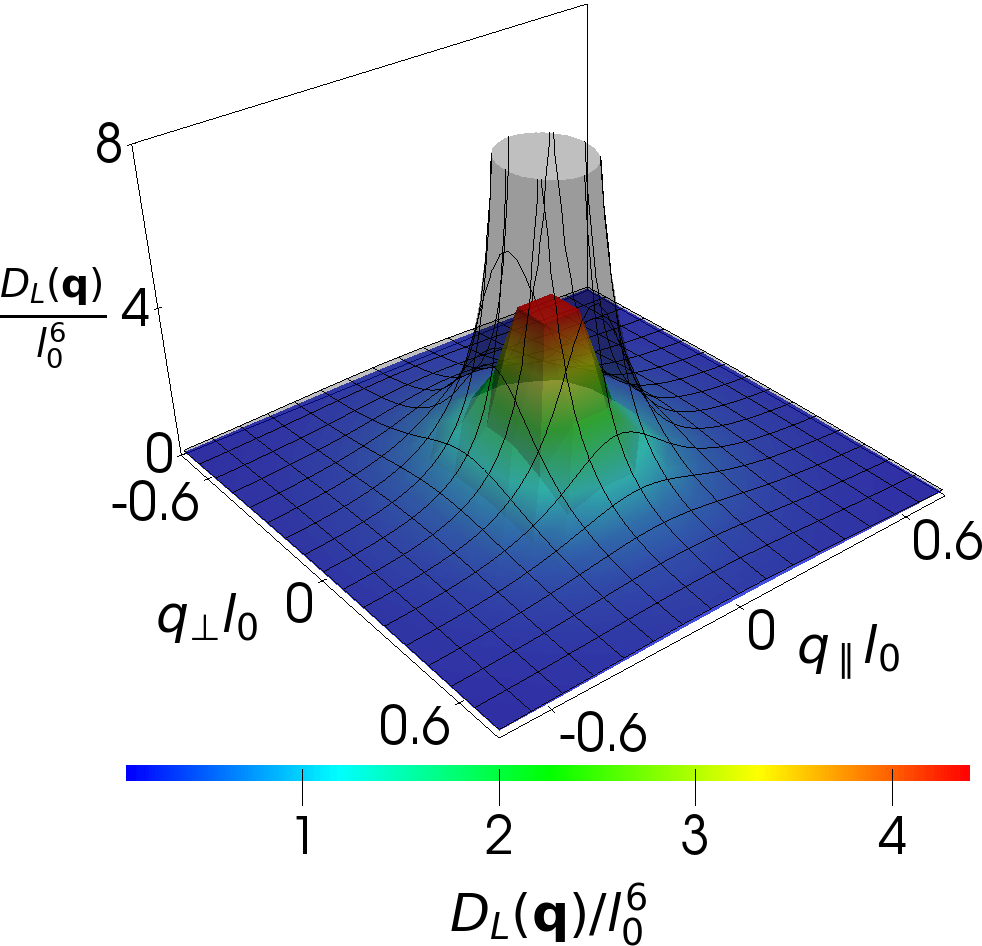}
	\caption{Longitudinal director fluctuations $D_L({q}_\perp l_0,q_\parallel l_0)$ calculated in simulations (solid), fitted (wireframe) by Eq.~(\ref{D_L}), for stiff (left) and flexible (right) chains of length $128l_0$.
}
\label{fig:D_L}
\end{center}
\end{figure}

In the same manner, we compute \cite{kostas_soft} also the longitudinal director fluctuation $D_L({\bf q}) = \langle \delta n_L({\bf q})\delta n_L(-{\bf q})\rangle/N$, with the theoretical expression \cite{nelson,svensek_jcp} 
\begin{equation}
	D_L({\bf q}) = {k_{\rm B} T\over \rho_0}{q_\parallel^2+B/\tilde G\over B q_\perp^2+\left(B/\tilde G+q_\parallel^2\right)
    		\left(K_1 q_\perp^2+K_3 q_\parallel^2\right)}
	\label{D_L}
\end{equation}
following  from Eq.~(\ref{f_q}).
For stiff chains, where the constraint Eq.~(\ref{recovered}) is expectedly strong, $D_L$ shows a characteristic strengthening \cite{nelson,kostas_soft} of the effective splay ($K_1$) elastic constant, Fig.~\ref{fig:D_L}.

The computed $S({q}_{\perp},q_{\parallel})$ and $D_L({q}_{\perp},q_{\parallel})$ landscapes, Figs.~\ref{fig:S(q)} and \ref{fig:D_L}, are fitted with 
Eqs.~(\ref{Svect}) and (\ref{D_L}) to extract the parameters $B$, $\tilde{G}$, $K_1$, $K_3$. Fig.~\ref{fig:S(q)} (bottom) shows cross sections of the two-dimensional structure factor fits. 
For small wave vectors it is verified that $S^{-1}(0,q_\parallel)$ is parabolic, while $S^{-1}(q_\perp,0)$ is essentially constant, as predicted by Eq.~(\ref{Svect}). 
The kinks at $q_\parallel l_0\approx \pm 0.5$ are attributed to microscopic effects not captured by the macroscopic theory, e.g. enhanced correlations within single chains or groups of neighboring chains \cite{binder}.

\begin{figure}[h]
\begin{center}	
	\includegraphics[width=75mm]{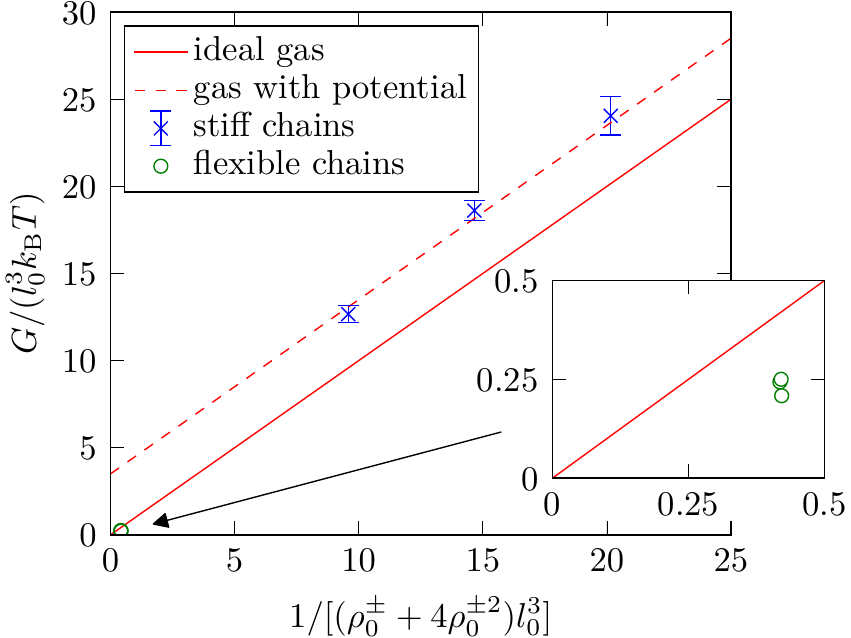} 
	\caption{Dimensionless strength of the constraint $G=\tilde G/(\rho_0 l_0 q_0)^2$, determined from the fits of the MC structure factor landscapes Fig.~\ref{fig:S(q)}, versus the dimensionless inverse density of the combined sources (solid line, no fitting parameter). Following Eq.~(\ref{fA}), an offset (dashed line) is fitted to the three points representing the stiff chains with $N_{\rm s}=\{32,64,128\}$.}
\label{fig:G}
\end{center}
\end{figure}

Using the averaged values $\rho_0$ and $a_0^r$, the strength $G$ of the constraint is determined from the fitting parameter $\tilde G$ and is plotted in dimensionless form in Fig.~\ref{fig:G} as a function of the dimensionless inverse density of chain ends/backfolds as suggested by Eq.~(\ref{f_avg}).
The average numbers of backfolds per chain are $\{0.33, 0.58, 1.1\}$ and $\{12, 24, 48\}$ for stiff and flexible chains with $N_{\rm s}=\{32, 64, 128\}$, respectively.
The direct comparison with the theoretical line not involving any fitting parameter
confirms the relevance of the prediction Eq.~(\ref{f_avg}). Especially the slopes agree remarkably. Moreover, the points corresponding to the flexible chains in Fig.~\ref{fig:G} (inset) show a highly reduced splay--density coupling, thus confirming the concept of the recovered polar order and the applicability of the conservation law Eq.~(\ref{recovered}) formulated on its basis, as well as the role of backfolds as sources in this conservation law.

\begin{figure}[h]
\begin{center}	
	\includegraphics[width=70mm]{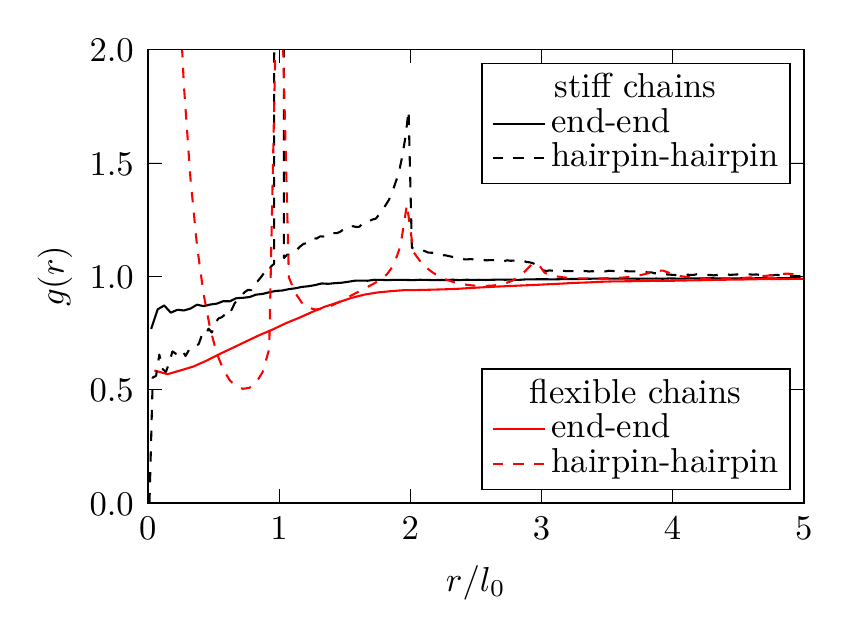} 
	\caption{RDFs of chain ends and backfolds (hairpins) for stiff (black) and flexible (red) chains with $N_{\rm s} = 128$ obtained in simulations. The end--hairpin RDFs (not shown) are qualitatively similar to the hairpin--hairpin RDFs.}
\label{fig:rdf}
\end{center}
\end{figure}

It is hard to overlook the hinted offset of the simulated stiff chain points from the theoretical solid line in Fig.~\ref{fig:G}.
We interpret it as a deviation from the noninteracting gas idealization of the chain ends/backfolds ($\equiv$ particles).
In fact, the computed end--end, hairpin--hairpin and end--hairpin radial distribution functions (RDFs), Fig.~\ref{fig:rdf}, show deviations of various kinds from the ideal gas behavior $g(r) = 1$ for $r\lessapprox 3l_0$. 
The end--end RDFs manifest simple repulsion. In contrast, the hairpin--hairpin RDFs have a complex structure due to contributions from hairpins on the same chain: small distances between sequential backfolds along the chain can assume only specific values, which explains the pronounced spikes (even for the stiff WLCs). Such 
small-scale effects, as well as distinguishing between intra- and intermolecular 
backfolds in Eq.~(\ref{f_avg}), are beyond the scope of the present macroscopic theory. 
It is however sensible to capture the interactions between the particles by an effective free-energy density $f(\rho^{s\pm})$ 
of the particle distribution $\rho^{s\pm}$,
\begin{equation}
	f(\rho^{s\pm}) = k_{\rm B} T \rho^{s\pm} \ln{\rho^{s\pm}\over\rho^{s\pm}_1} + {1\over 2}A(\rho^{s\pm})^2,
\end{equation}
where the pair-interaction free-energy density is proportional to $(\rho^{s\pm})^2$ by definition, while all details of this interaction are contained in a phenomenological second virial coefficient $A$; $A>0$ stands for an effective repulsion and vice versa.
The insignificant constant $\rho_1^{s\pm}$ is determined by fixing the equilibrium density $\rho^{s\pm}_0$, i.e., $f'(\rho^{s\pm}_0) = 0$. In the next, quadratic order of $\Delta\rho^{s\pm} = \rho^{s\pm}-\rho^{s\pm}_0$, we have
\begin{equation}
	f(\Delta\rho^{s\pm}) = {1\over 2}\left({k_{\rm B} T\over\rho_0^{s\pm}}+A\right)(\Delta\rho^{s\pm})^2,
    \label{fA}
\end{equation}
which could explain the rather constant positive offset ($A>0$) of $G(1/\rho^{s\pm}_0)$ for the set of stiff chains in Fig.~\ref{fig:G} (dashed line). This repulsive effective interaction makes nonequilibrium excursions $\Delta\rho^{s\pm}$ more expensive, Eq.~(\ref{fA}), and hence the constraint Eq.~(\ref{recovered}) is stronger.
Note however that the strength $A$ of the effective interaction depends on the composition of the particle gas, i.e., the ratio $\rho_0^{2\pm}/\rho_0^\pm$, and furthermore that also the RDFs depend on $N_{\rm s}$ and other parameters.

In conclusion, we have established a consistent macroscopic description of the splay--density coupling in semiflexible main-chain nematic polymers with hairpins, using a vectorial continuity constraint for the recovered polar order of chain tangents and introducing chain backfolds as its new type of sources besides chain ends. In the minimal spirit, we unified both types of sources to a mixture of two ideal gases with fixed composition. Conducting detailed Monte Carlo simulations of nematic monodomain melts of worm-like chains with variable length and flexibility, we demonstrated via their structure factors that the chain backfolding weakens the splay--density coupling, and showed
how this weakening can be quantified on the macroscopic level.


\begin{acknowledgments}
Acknowledged is the support of the Slovenian Research Agency---Grants P1-0002, P1-0055, J1-7435, J1-7441 (A.P., D.S., R.P., M.P.).
\end{acknowledgments}

\end{document}